\documentclass[prl, aps, 10pt, showkeys, twocolumn, superscriptaddress]{revtex4-1}

\usepackage[english]{babel}
\usepackage{amsmath}
\usepackage{amssymb}
\usepackage{graphicx}
\usepackage{hyperref}

    \def\a{\alpha}
    \def\b{\beta}
    
    \def\<{\langle}
    \def\>{\rangle}

    \def\vphi{\varphi}
    \def\pt{\partial}

  \newcommand{\eq}[1]{
    \begin{equation}
    {#1}
    \end{equation}}

    \newcommand{\mq}[1]{
    \begin{equation}
    \begin{split}
    #1
    \end{split}
    \end{equation}}

    \newcommand{\nmq}[1]{
    \begin{multline}
    #1
    \end{multline}}

    \newcommand{\mtwo}[4]{
    \begin{pmatrix}
    #1 & #2 \\ #3 & #4
    \end{pmatrix}
    }

\begin{document}

\title{Non-equilibrium noise in transport across a tunneling contact between $\nu = 2/3$ fractional quantum Hall edges}

\author{O.~Shtanko}
\affiliation{Department of Physics, Massachusetts Institute of Technology, Cambridge MA 02139, USA}
\affiliation{Physics Department, Taras Shevchenko National University of Kyiv, Kyiv, 03022, Ukraine}
\author{K.~Snizhko}
\affiliation{Physics Department, Taras Shevchenko National University of Kyiv, Kyiv, 03022, Ukraine}
\affiliation{Department of Physics, Lancaster University, Lancaster, LA1~4YB, UK}
\author{V.~Cheianov}
\affiliation{Department of Physics, Lancaster University, Lancaster, LA1~4YB, UK}

\begin{abstract}
  In a recent experimental paper [A. Bid et al., Nature \textbf{466}, 585 (2010)] a qualitative confirmation of the existence of upstream neutral modes at the $\nu = 2/3$ quantum Hall edge was reported. Using the chiral Luttinger liquid theory of the quantum Hall edge we develop a quantitative model of the experiment of Bid et al. A good quantitative agreement of our theory with the experimental data reinforces the conclusion of the existence of the upstream neutral mode. Our model also enables us to extract important quantitative information about non-equilibrium processes in Ohmic and tunneling contacts from the experimental data. In particular, for $\nu = 2/3$, we find a power-law dependence of the neutral mode temperature on the charge current injected from the Ohmic contact.
\end{abstract}

\maketitle

\section{Introduction}

The quasi-one-dimensional edge channels supported by fractional quantum Hall (FQH) states have for a long time attracted attention of both theorists and experimentalists. During the 1980s models emphasizing role of edge states for FQH transport developed into a powerful field-theoretical framework of a chiral Luttinger liquid (CLL) \cite{WenReview}. A very rigid mathematical structure of the latter led to a number of nontrivial predictions such as fractionally charged quasiparticles, and excitations with anyonic or even non-Abelian statistics. Some of these predictions have been tested experimentally while others still pose a challenge to experimentalists.

One of the milestones in the experimental studies of the FQH effect is the recently reported observation \cite{HeiblumExp} of the neutral current (a transport channel which does not carry electric charge) at the edge of the $\nu = 2/3$ FQH state. The $\nu = 2/3$ state is one of the simplest for which the CLL theory is not consistent without a neutral current. Moreover, the predicted flow direction of this current is opposite to the electrons' drift velocity \cite{KaneFisher, KaneFisherPolchinski} and thus contradicts intuition based on the magnetic hydrodynamics \cite{AleinerGlazman}.

Apart from the detection of the upstream neutral mode, the design of the experiment, Ref.~\cite{HeiblumExp}, gave access to a significant amount of quantitative data characterizing the system \cite{HeiblumExp, HeiblumExp2}. This motivates the present work, where a detailed quantitative description of the experiment is developed basing on the minimal $\nu = 2/3$ edge model worked out in \cite{KaneFisher, KaneFisherPolchinski} and supported by numerical simulations of small systems \cite{HuRezayi23EdgeMicroscopicStudy, WuJain23EdgeMicroscopicStudy}. Within the developed framework we analyze the data of \cite{HeiblumExp} in order to (a) check its consistence with the minimal $\nu = 2/3$ edge model quantitatively and (b) extract new information about $\nu = 2/3$ edge physics.

While we find excellent agreement of our theory with the experimental data of Ref.~\cite{HeiblumExp}, we would like to remark that a number of alternative theories have been proposed recently in order to explain other experimental results such as those of Ref.~\cite{HeiblumForFerraro}. These theories extend the minimal $\nu=2/3$ edge model by introducing new physics, such as edge reconstruction \cite{MeirGefen23EdgeReconstruction} or bandwidth cutoffs \cite{Ferraro}, at some intermediate energy scale. Such extensions can be incorporated into our framework. However, as they contain additional unknown parameters, their comparison against experiment can only be insightful with more independent experimental data.

\section{Description of the experiment \cite{HeiblumExp}}

Here we briefly discuss the experiment \cite{HeiblumExp} where the upstream neutral currents at the quantum Hall (QH) edge were investigated.

\begin{figure}[tbp]
  \center{\includegraphics[width=1\linewidth]{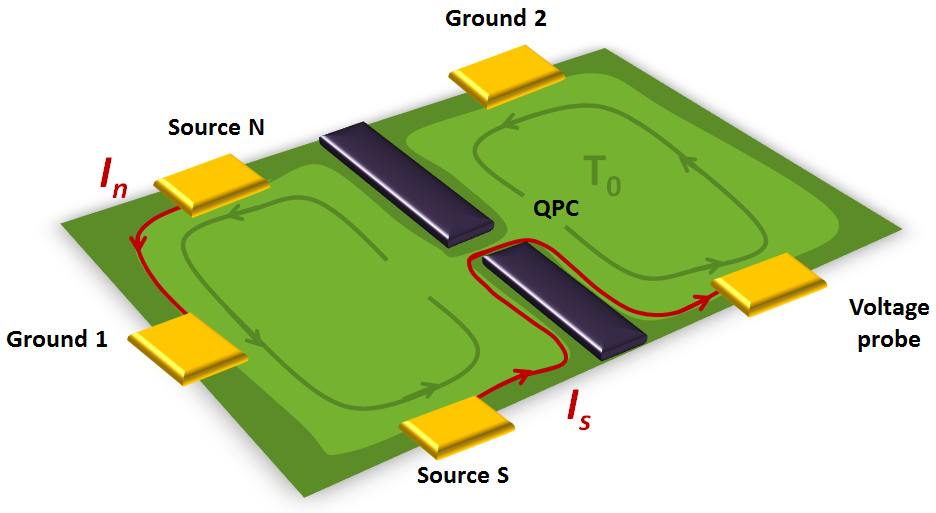}}
  \caption{\textbf{Scheme of the experimental device}. (Color online). Contacts \textit{Ground 1} and \textit{Ground 2} are grounded. \textit{Source N} and \textit{Source S} are used to inject some electric current into the system. Measurement of the electric current and its noise is performed at the \textit{Voltage probe}.}
  \label{fig_exp_setup}
\end{figure}

Figure \ref{fig_exp_setup} shows a sketch of the experimental device. The green (color online) region is the AlGaAs heterostructure with the light-green showing where the 2DEG (two-dimensional electron gas) is actually present. The sample is in the transverse magnetic field so that the filling factor is 2/3 and the corresponding quantization of the Hall conductivity is observed. Green arrows show the direction of the electrons' drift velocity which coincides with the flow direction of the charge transporting channel (charged mode). Yellow patches represent Ohmic contacts. The purple rectangular pads on top of the sample are the gates which allow one to make a constriction which plays the role of tunneling junction (denoted as \textit{QPC}~--- quantum point contact --- in the figure). Contacts \textit{Ground 1} and \textit{Ground 2} are grounded. \textit{Source N} and \textit{Source S} are used to inject electric current into the device. Measurements of electric current and its noise are performed at \textit{Voltage probe}.

The idea of the experiment is as follows. Suppose a current $I_n$ is injected into \textit{Source N}. If the edge supports only one chirality (counterclockwise) then  anything injected into \textit{Source N} will be absorbed by \textit{Ground 1} and have no effect on \textit{Voltage probe}. However, if we assume that there is a neutral mode flowing clockwise, information about the events in \textit{Source N} carried by the neutral mode may reach \textit{QPC}. In \textit{QPC} such information may be transmitted to the opposite edge and then transported to \textit{Voltage probe} by the charged mode. In particular, let us assume that the injection of the current $I_n$ excites the neutral mode flowing out of \textit{Source N} towards \textit{QPC}. Due to the tunneling across \textit{QPC} of quasiparticles having both charged and neutral degrees of freedom the neutral mode excitations will be converted into the current noise at \textit{Voltage probe}. Thus, the presence of the counterpropagating neutral mode implies that the noise observed at \textit{Voltage probe} should depend on the current $I_n$. Such a dependence was reported in \cite{HeiblumExp}.

The observation of the theoretically predicted upstream neutral mode is a very important qualitative result. However, experimental techniques and numerical data reported in \cite{HeiblumExp} go far beyond this achievement providing a lot of implicit quantitative information about current fractionalization in Ohmic contacts, transport along the QH edges and quasiparticle tunneling across the \textit{QPC}. In order to effectively utilize this quantitative information one needs an analytical theory of the experiment based on the modern understanding of the FQH edge. The goal of this work is to discuss the results of \cite{HeiblumExp} within such a theoretical framework.

\section{Theoretical picture of the experiment}

Our theoretical description of the experiment has three key ingredients: the effective theory of the quantum Hall edge, a model of the \textit{QPC}, and phenomenological assumptions about the interaction of the Ohmic contacts with the QH edge. The former two are based on the standard theoretical framework which we briefly review in the next section. In this section we focus on the general picture of the experiment, paying particular attention to the assumptions regarding the Ohmic contacts.

\begin{figure}[tbp]
\center{\includegraphics[width=1\linewidth]{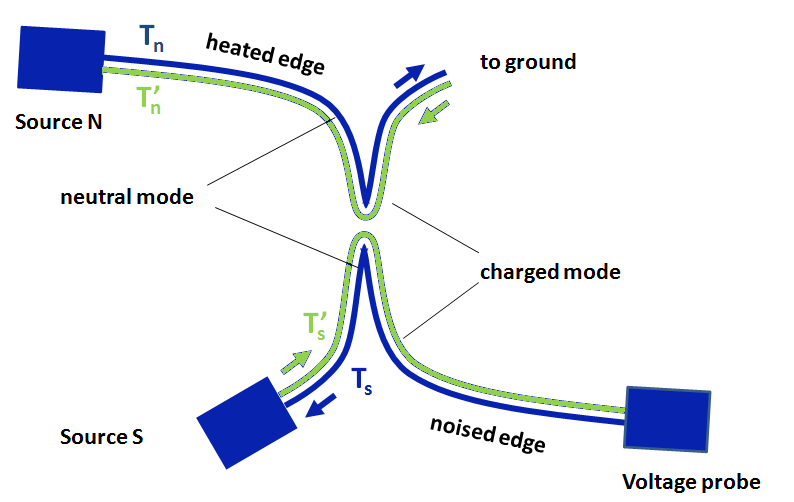}}
\caption{\textbf{Theoretical picture of the experiment}. (Color online). The injected current $I_n$ "heats" the neutral mode of the upper edge to the temperature $T_n$. Equilibration processes between the charged and the neutral modes lead to the charged mode temperature $T_n' = T_n$. Both modes at the lower edge have the temperature of the environment: $T_s' = T_s = T_0$. Tunneling of the quasiparticles at the constriction induces extra noise in the charged mode of the lower edge which is detected at the \textit{Voltage probe}. Injection of the current $I_s$ only changes the chemical potential of the charged mode of the lower edge.}
\label{fig_exp_model}
\end{figure}

Our theoretical model of the experiment is illustrated in Fig.~\ref{fig_exp_model}. Each edge supports one counterclockwise charged mode and one clockwise neutral mode. The two edges approach each other in the \textit{QPC} region where the tunneling of the quasiparticles between the edges occurs. Our quantitative theory is developed for the case of weak quasiparticle tunneling. The Ohmic contacts are shown as rectangles. We assume that any excitations of neutral and charged modes are fully absorbed by the Ohmic contacts they flow into. We further assume strong equilibration mechanisms at the edge so that the hydrodynamic description can be used. That is, each edge can be characterized by local point-dependent thermodynamic variables including the charged mode chemical potential $\mu^{(c)}$, the charged mode temperature $T'$ and the neutral mode temperature $T$, and any other thermodynamic variables arising due to the existence of extra conserved quantities. We assume that in the absence of currents ($I_n = 0$ and $I_s = 0$) the edges are in equilibrium with the environment so that all modes' temperatures are equal to the base temperature $T_0$ and the chemical potentials are equal to zero. Away from this state the temperatures and chemical potentials are unknown functions of $I_n$ and $I_s$, and other thermodynamic variables are assumed to be unaffected by the injection of currents $I_n$, $I_s$. The functions $\mu^{(c)}(I_n, I_s)$, $T(I_n, I_s)$, $T'(I_n, I_s)$ for each edge are defined by the interaction of the Ohmic contact with the edge, however no predictive theoretical model of such interaction is known today. As we show, these functions can be inferred from the experimental data under some plausible phenomenological assumptions.

We assume that there is a strong heat exchange between the modes at each edge. In this approximation the local temperatures of the two modes coincide at each point along the edge: $T_n = T_n'$, $T_s = T_s'$. Moreover, following \cite{HeiblumExp} we assume that the lower edge temperature is equal to the base temperature ($T_s = T_s' = T_0$); that is, the electric current $I_s$ injected by the Ohmic contact \textit{Source S} does not induce any nonequilibrium noise to the lower edge charged mode.

\section{Formalism of the edge field theory}

In this section we give a brief overview of the CLL formalism \cite{WenReview, WenCLL, FrohlichAnomaly} which is believed to provide the effective theoretical description of a fractional QH edge. We then focus on a particular edge model relevant to the experiment \cite{HeiblumExp}. We conclude this section by a discussion of the model Hamiltonian describing the tunneling of quasiparticles between the QH edges.

\subsection{General formalism\footnote{In this and the following sections we put $e = \hbar = k_B = 1$ unless the opposite is stated explicitly. Here $e$ is the elementary charge, $\hbar$ is the Planck constant, $k_B$ is the Boltzmann constant.}}

Abelian QH edge theories are usually formulated in terms of bosonic fields $\vphi_i(x, t)$, where $t$ is time and $x$ is the spatial coordinate along the edge. Each field $\vphi_i$ represents an edge mode. Suppose that we have $N$ edge modes and correspondingly $N$ fields $\vphi_i$ with $i = 1,...,N$. Then the dynamics of the edge is described by the effective action\footnote{In fact, the action \eqref{action} has to be used with care because its chiral nature imposes implicit constraints on the external perturbation $a_{\mu}$. This problem does not emerge in Hamiltonian formalism used in \cite{WenCLL}.}
\nmq{\label{action}
S = \frac1{4\pi} \int dx dt \sum_i \Bigl( -\chi_i D_x\vphi_i D_t \vphi_i-v_i(D_x\vphi_i)^2 +\\
 +q_i \varepsilon^{\mu\nu} a_\mu \pt_\nu \vphi_i\Bigl),
}
where $v_i\in\mathbb{R}_+$ are the propagation velocities, $\chi_i = \pm1$ represent chiralities of the modes (plus for the clockwise and minus for the counterclockwise direction), and $a_\mu(x,t)$ is the electromagnetic field potential at the edge. Covariant derivatives are defined as $D_\mu\vphi_i = \pt_\mu\vphi_i-\chi_i q_i a_\mu$. The coupling constants $q_i$ provide information on how the electric charge is distributed between the modes. The symbol $\varepsilon^{\mu\nu}$ denotes the fully antisymmetric tensor with $\mu, \nu$ taking values $t$ and $x$ (or 0 and 1 respectively) and $\varepsilon^{t x} = \varepsilon^{0 1} = 1$.

Conservation of total electric current in the whole volume of a 2D sample leads to the condition \cite{WenCLL, FrohlichAnomaly}
\eq{\label{anomaly_cancellation}\sum_i \chi_i q_i^2 = \nu.}
The electric current at the edge is
\eq{\label{el_current operator} J^\mu = \frac{\delta S}{\delta a_\mu} = \frac{1}{2\pi}\sum_i q_i\varepsilon^{\mu\nu}D_\nu\vphi_i + \frac{\nu}{4\pi}\varepsilon^{\mu\nu}a_\nu.}
In the presence of the electric field it is not conserved:
\eq{\pt_\mu J^\mu = - \frac{\nu}{4\pi}\varepsilon^{\mu\nu}\pt_\mu a_\nu\neq0,}
which is a manifestation of the inflow of the Hall current from the bulk.

In the absence of the electric field $a_\mu(x,t) = 0$ the current is conserved and has the form
\eq{\label{el_current operator_2} J^\mu = \frac{1}{2\pi}\sum_i q_i\varepsilon^{\mu\nu}\pt_\nu\vphi_i, \qquad \pt_\mu J^\mu = 0}

In the rest of this section we assume that $a_\mu(x,t) = 0$.

Beyond the electric current one can also define neutral currents
\eq{\label{neut_current operator} J^{\mu}_{\mathrm{n}} = \frac{1}{2\pi}\sum_i p_i\varepsilon^{\mu\nu}\pt_\nu\vphi_i, \qquad \pt_\mu J^{\mu}_{\mathrm{n}} = 0,}
with vector $\mathbf{p} = (p_1,...,p_N)$ being linearly independent of vector $\mathbf{q} = (q_1,...,q_N)$.\footnote{The conserved neutral currents can give rise to neutral modes' chemical potentials $\mu^{(n)}$~--- thermodynamic quantities dual to the neutral charges. In the main text, as we pointed out in the previous section, we assume that these neutral chemical potentials are not involved in the experiment we are going to analyze. However, for the sake of generality we include them in formulas in Appendices~A, E.}

The quantized fields $\vphi_i$ can be presented as follows
\begin{multline}
\label{free_boson_solution}\vphi_i(x,t) = \vphi^0_i + \frac{2 \pi}{L} \pi^0_i X_i + \\
 +i \sum_{n = 1}^{\infty} \sqrt{\frac{2\pi}{L k}} \Bigl( a_i(k)\exp(-i k X_i) - a^\dag_i(k)\exp(i k X_i)\Bigl)
\end{multline}
where $X_i = -\chi_i x+v_i t$, $k = 2 \pi n/L$, $n \in \mathbb{N}$; $L \rightarrow \infty$ is the system size, $a_i(k)$ and $a^\dag_i(k)$ are the annihilation and the creation operators respectively, and $\vphi_0$ and $\pi_0$ are the zero modes:
\eq{[a_i(k),a_j^\dag(k')]= \delta_{ij} \delta_{k k'}, \qquad [\pi^0_i,\vphi^0_i]=-i \delta_{ij}.}

The fields $\vphi_i$ obey the commutation relation of chiral bosons:
\eq{\label{comm_relations}[\vphi_i(x,t),\vphi_j(x',t')] = -i \pi \mathrm{sgn}(X_i-X_i')\,\delta_{ij}.}

The edge supports quasiparticles of the form
\eq{\label{vertex_quasi}V_{\mathbf{g}}(x,t) = \left( \frac{L}{2\pi} \right)^{-\sum_i g_i^2/2} :\exp\left(i\sum_i g_i\vphi_i(x,t)\right):,}
which are important for the processes of tunneling at the \textit{QPC}. The notation $: ... :$ stands for the normal ordering, $\mathbf{g} = (g_1,...,g_N)$, and $g_i\in\mathbb{R}$ are the quasiparticle quantum numbers.

Among the quasiparticle fields there has to be a field representing an electron which is the fundamental constituent particle:
\mq{\label{electron_operator}\psi(x,t) = \left( \frac{L}{2\pi} \right)^{-\sum_i a_i^2/2} :\exp\left(i\sum_i a_i \varphi_i(x,t)\right):,}
$a_i\in\mathbb{R}$. Minimal models of the QH states of Jain series $\nu = N/(2N \pm 1)$ have $N$ electron operators each representing a composite fermion Landau level:
\nmq{\label{electron_operators}\psi_{\alpha}(x,t) = \left( \frac{L}{2\pi} \right)^{-\sum_i e_{\alpha i}^2/2} :\exp\left(i\sum_i e_{\alpha i} \varphi_i(x,t)\right):,\\ e_{\alpha i}\in\mathbb{R}.}

The electron fields have to satisfy the following constraints:
\mq{&\{\psi_{\alpha}(x,t),\psi_{\alpha}(x',t)\} = 0 \\
&\psi_{\alpha}(x,t)\psi_{\beta}(x',t) \pm \psi_{\beta}(x',t)\psi_{\alpha}(x,t)=0,\quad \alpha\neq\beta \\
 &[J^{0}(x,t),\psi_{\alpha}(x',t)] = \delta(x-x')\psi_{\alpha}(x,t).}
where $J^0$ is charge density operator defined in Eq.~\eqref{el_current operator_2}, $\{...\}$ denotes the anti-commutator, and a plus or minus sign in the second equation can be chosen independently for each pair $(\alpha, \beta)$; $\alpha, \beta = 1,...,N$.

For the parameters $e_{\alpha i}$ in Eq.~\eqref{electron_operators} these constraints imply
\eq{\label{eq_for_e_basis} \mathbf{e}_{\alpha}\cdot \mathbf{e}_{\alpha}\in2\mathbb{Z}+1, \quad \mathbf{e}_{\alpha}\cdot \mathbf{e}_{\beta}\in\mathbb{Z}, \quad \mathbf{q}\cdot \mathbf{e}_{\alpha} = -1}
where we defined $\mathbf{e}_{\alpha} = (e_{\alpha 1},...,e_{\alpha N})$ and $\mathbf{q} = (q_1,...,q_N)$ with $q_i$ being the coupling constants from the action \eqref{action}, and the operation $\mathbf{A}\cdot \mathbf{B} \equiv \sum_{i=1}^{N} \chi_i A_i B_i$.

Equations \eqref{eq_for_e_basis} have many inequivalent solutions each defining a topological QH class. It is convenient to parametrize these classes with the help of the K-matrix:
\eq{K_{\a\b} = \mathbf{e}_\a\cdot \mathbf{e}_\b.}

Consider now a QH fluid corresponding to a particular solution $\{\mathbf{e}_1,...,\mathbf{e}_N\}$ of Eqs.~\eqref{eq_for_e_basis}. The spectrum of the quasiparticles \eqref{vertex_quasi} present in the model is determined from the requirement of mutual locality with all the electron operators:
\eq{\label{elect_quasy} \psi_{\alpha}(x,t)V_g(x',t) + s V_g(x',t)\psi_{\alpha}(x,t)=0,}
where $s$ is either $+1$ or $-1$ depending on the particular quasiparticles.

This leads to the following restrictions on the parameters $g_i$ in Eq.~\eqref{vertex_quasi}:
\eq{\label{qp_spectrum_restrictions}\mathbf{g}\cdot \mathbf{e}_{\alpha} = n_{\alpha} \in \mathbb{Z}, \quad \alpha = 1,...,N,}
where $\mathbf{g} = (g_1,...,g_N)$. The set of numbers $n_{\alpha}$ completely defines the properties of a quasiparticle operator.

For the following considerations two quantum numbers of the quasiparticle operator \eqref{vertex_quasi} are of particular importance: the electric charge $Q$ and the scaling dimension $\delta$. They are given by
\begin{eqnarray}
\label{Charge}
Q(\mathbf{n}) &=& \mathbf{q}\cdot \mathbf{g} = \sum_{\a\b} K_{\a\b}^{-1}n_\b,\\
\label{ScalDim}
\delta(\mathbf{n}) &=& \frac 12\sum_i g_i^2.
\end{eqnarray}

\subsection{The minimal model the $\nu = 2/3$ QH edge}

Here we use the general principles discussed above to obtain the minimal model of the $\nu = 2/3$ QH edge. This model emerges from different semi-phenomenological theoretical approaches to the QH edge \cite{WenAntiChiral23, MacDonaldAntiChiral23} and is the most likely candidate to describe this fraction \cite{WuJain23EdgeMicroscopicStudy}.

First, we note that it is impossible to satisfy the constraints \eqref{anomaly_cancellation} and \eqref{eq_for_e_basis} assuming that $N = 1$. For $N = 2$ we choose the charge vector\footnote{Note that there exists an infinite freedom in the choice of the vector $\mathbf{q}$ giving rise to infinitely many physically inequivalent theories. However, as it was shown in \cite{KaneFisher, KaneFisherPolchinski} by perturbative RG analysis, the choice \eqref{charge_vector} leads to a theory stable against disorder scattering.}
\eq{\label{charge_vector}\mathbf{q} = (\sqrt{2/3},0)}
and chiralities
\eq{\chi_1 = 1,\quad \chi_2 = -1.}

Then equations \eqref{eq_for_e_basis} lead to an infinite one-parameter family of solutions:
\begin{eqnarray}
\mathbf{e}_1 = \left(- \sqrt{\frac32},\,\sqrt{\frac32+2m+1}\right),\\
\mathbf{e}_2 = \left(- \sqrt{\frac32},\,-\sqrt{\frac32+2m+1}\right),
\end{eqnarray}
where $m=-1,0,1,2,...$

The electron operators have the smallest scaling dimension for $m = -1$, which gives
\eq{\mathbf{e}_{1,2} = \left(- \sqrt{\frac32},\,\pm\sqrt{\frac12}\right)}
and the K-matrix
\eq{\label{K_min}K = \mtwo{1}{2}{2}{1}.}
This defines the minimal model of the $\nu = 2/3$ QH edge.

The quasiparticle spectrum of the model is defined by Eq.~\eqref{qp_spectrum_restrictions}. The parameters of the three excitations which are most relevant for tunneling across the \textit{QPC} are given in Table~\ref{tab:exc}.

\begin{table}[h!]
  \caption{\label{tab:exc} Parameters of the most relevant excitations in the minimal model of the $\nu = 2/3$ QH edge (see Eqs. \eqref{vertex_quasi}, \eqref{qp_spectrum_restrictions}, \eqref{Charge}, and \eqref{ScalDim}).}
  \begin{center}
  \begin{tabular}{|c|c|c|c|c|}
  \hline
  Type & $\qquad g_1\qquad$ & $\qquad g_2\qquad$ & $\quad Q\quad$ & $\quad\delta\quad$ \\
  \hline
  1 &  $\sqrt{1/6}$& $\sqrt{1/2}$&$1/3$&$1/3$ \\
  \hline
  2 &  $\sqrt{1/6}$& $-\sqrt{1/2}$&$1/3$&$1/3$ \\
  \hline
  3 &  $\sqrt{2/3}$& $0$&$2/3$&$1/3$ \\
  \hline
\end{tabular}
\end{center}
\end{table}

We find it convenient to define the neutral current \eqref{neut_current operator} with
\eq{\label{neut_charge_vector}\mathbf{p} = (0, -1).}

\subsection{Tunneling of quasiparticles across the \textit{QPC}}

Wherever the two QH edges approach each other at a distance on the order of the magnetic length processes of quasiparticle exchange between the edges are possible. It is widely accepted \cite{WenCLL, Saleur, KaneFisherTunnHam} that such processes can be described by adding the following term to the Hamiltonian:
\eq{\label{tun_ham} H_{T} = \sum_{\mathbf{g}} \eta_{\mathbf{g}} V_{\mathbf{g}}^{(u)\dag}(0,t)V^{(l)}_{\mathbf{g}}(0,t) + \mathrm{h.c.},}
where the superscripts $(u), (l)$ label quantities relating to the upper and the lower edge respectively; for simplicity we assume that tunneling occurs at the origin. In the case of weak tunneling across the bulk of the QH state the sum runs over all quasiparticles in the model. However, at small energies quasiparticles with the smallest scaling dimension $\delta(\mathbf{g})$ have the largest tunneling amplitude $\eta_{\mathbf{g}}$, thus giving the most important contribution.

\section{Calculation of observable quantities}

In this section we derive analytical expressions for two observable quantities as functions of the experimentally variable parameters. These quantities include the tunneling rate that is the ratio of the current tunneling across the \textit{QPC} to the \textit{Source S} current $I_s$ and the excess noise in the \textit{Voltage probe} which is the noise in the \textit{Voltage probe} in the presence of currents $I_n$, $I_s$ less the equilibrium noise at $I_n = I_s = 0$. We further demonstrate that it is advantageous to consider the ratio of these quantities rather than each separate one. This way the influence of non-universal physics of the tunneling contact can be reduced.

Our expressions for the excess noise and the tunneling rate, presented in Eqs.~\eqref{noise_expr}-\eqref{j_s}, are in full agreement with Eqs.~(10) and (11) of Ref.~\cite{Rosenow}.

\subsection{Tunneling rate}

As it was mentioned in the previous section, the most important contribution to the tunneling processes is due to the most relevant excitations. Such excitations are listed in Table~\ref{tab:exc}, and we restrict our considerations to these excitations only. To this end we introduce the following notation $\psi_i(x,t) = V_{\mathbf{g}_i}(x,t)$ where $\mathbf{g}_i$, $i = 1,2,3$ are the three most relevant quasiparticle vectors given in Table~\ref{tab:exc}.

The tunneling Hamiltonian can then be written as
\begin{eqnarray}
\label{tun_ham_rel} H_{T} &=& \sum_{i = 1}^{3} \eta_i A_i(t) +\eta^*_i A^\dag_i(t),\\
\label{tun_oppsA} A_i(t) &=& \psi_i^{(u)\dag}(0,t)\psi^{(l)}_i(0,t)
\end{eqnarray}
where the superscripts $(u), (l)$ label quantities relating to the upper and the lower edge respectively and $\eta_i$ are unknown complex phenomenological parameters.

We calculate the tunneling current within the second-order perturbation theory in the tunneling Hamiltonian. The detailed derivation can be found in Appendix B. The resulting tunneling rate is given by the Kubo formula:
\eq{\label{tun_general}r = \left|\frac{I_{T}}{I_s}\right|= \left| \frac{1}{I_s}\sum_{i} |\eta_i|^2  Q_i\int\limits_{-\infty}^{\infty} d\tau \left\langle \left[A_i(\tau), A_i^\dag(0)\right]\right\rangle \right|,}
where $I_{T}$ is the tunneling current and $I_s$ is the current originating from \textit{Source S}. Apart from $r$, paper \cite{HeiblumExp} uses $t = 1-r$.

\subsection{Excess noise}

Noise spectral density of the electric current flowing into the \textit{Voltage probe} (see Fig.~\ref{fig_exp_model}) can be calculated as the Fourier transform of the two-point correlation function of the current operator~$I$,
\eq{\label{noise_general} S(\omega) = \int\limits_{-\infty}^\infty d\tau \exp\bigl(i\omega\tau \bigl) \frac 12 \Bigl\<\Bigl\{\Delta I(0),\Delta I(\tau)\Bigl\}\Bigl\>,}
where $\{\dots\}$ denotes the anti-commutator, and \mbox{$\Delta I = I - \<I\>$}.\footnote{We must note here that there are two conventions concerning the definition of the noise spectral density. While some authors (see, e.g., \cite{Lesovik1989}) use the same definition as we do, others (see, e.g., \cite{MartinLandauer}, \cite{HeiblumExp}) adopt the definition which is twice as large as ours. Thus our results must be multiplied by 2 in order to be compared with the data of \cite{HeiblumExp}.}

It is convenient to separate the operator $I$ of the full current flowing to the \textit{Voltage probe} into $I_0 + I_T$ with \mbox{$I_0 = J^{\mu (l)}(x = -0, t = 0)|_{\mu = 1 = x}$} being the spatial component of the operator $J^{\mu}(x,t)$ defined in Eq.~\eqref{el_current operator_2} which represents the electric current flowing along the lower edge just before the tunneling point, and $I_T$ being the tunneling current operator. Then the noise can be represented as follows:
\begin{eqnarray}
\label{Noise_sep_1}
S(\omega) &=& S_{00}(\omega) + S_{0T}(\omega) + S_{0T}(-\omega) + S_{TT}(\omega),\\
S_{ab}(\omega) &=& \int\limits_{-\infty}^\infty d\tau \exp\bigl(i\omega\tau \bigl) \frac 12 \Bigl\<\Bigl\{\Delta I_{a}(0),\Delta I_{b}(\tau)\Bigl\}\Bigl\>,
\label{Noise_sep_2}
\end{eqnarray}
with $\Delta I_{a} = I_a - \<I_a\>$, indices $a, b$ take values $0$ and $T$.

We are interested in the low-frequency component measured in the experiment. To a good approximation this can be replaced by the zero-frequency component $S(\omega = 0)$. Within the second order perturbation theory we find
\begin{eqnarray}
\label{noise_00}
S_{00}(0) &=& \frac{\nu}{2\pi} T_{s},\\
\label{noise_TT}
S_{TT}(0) &=& \sum_i |\eta_i|^2 Q_i^2 \int\limits_{-\infty}^\infty d\tau \bigl\langle \{A_i(0), A_i^{\dag}(\tau)\}\bigl\rangle,
\end{eqnarray}
\nmq{
\label{noise_0T}
S_{0T}(0) = \frac{1}{2} \sum_{i} |\eta_i|^2 Q_i \int\limits_{-\infty}^\infty d\tau \int\limits_{-\infty}^\infty d\tau' \\
\bigl\langle \{ \Delta I_0(0), [A_i(\tau'), A_i^{\dag}(\tau)] \} \bigl\rangle.
}
We remind the reader that $T_s$ is the lower edge temperature in the neighborhood of the \textit{QPC}. These formulas are derived in Appendix~C.

The contribution $S_{00}$ is the Johnson-Nyquist noise of the lower edge. If we restore $e$, $\hbar$, and $k_B$ we see that $S_{00}(0) = k_B T_s/R$, $R = 2 \pi \hbar / (\nu e^2) = h/(\nu e^2)$. Since the \textit{Voltage probe} contact not only absorbs the lower edge charged mode but also emits another charged mode which flows to the right of it, the actual Nyquist noise measured in the contact will be $S_{JN}(0) = 2 k_B T_s/R$, in agreement with general theory of Johnson-Nyquist noise. The factor of 2 difference from the Nyquist noise expression used in~\cite{HeiblumExp} is due to the noise spectral density definition as discussed in footnote~\cite{Note5}.

Following \cite{HeiblumExp} we define the excess noise
\eq{\tilde{S}(0) = S(0) - S_{\mathrm{eq}}(0),}
where $S_{\mathrm{eq}}$ is the equilibrium noise spectral density (i.e. the noise when $I_s = 0$ and $I_n = 0$, meaning that the edge temperatures are equal to the base temperature: $T_s = T_n = T_0$). It turns out that $S_{\mathrm{eq}}(0) = S_{00}(0)$ resulting in
\eq{\label{excess_noise}\tilde{S}(0) = 2 S_{0T}(0) + S_{TT}(0).}
This fact is proven in Appendix~D using the explicit formulas for $S_{0T}(0)$, $S_{TT}(0)$ obtained in Appendices C1 and C2.

\subsection{Noise to tunneling rate ratio\footnote{In this section we restore the elementary charge $e$, the Planck constant $\hbar = h/2\pi$, and the Boltzmann constant $k_B$ in order to simplify use of our formulas for comparison with experimental data.}}

Expressions \eqref{tun_general}, \eqref{noise_TT}, and \eqref{noise_0T} depend on the tunneling constants $\eta_i$. It is well known (see e.g. \cite{TunnellingRate0, TunnellingRate1, TunnellingRate2} and references therein) that the tunneling amplitudes $\eta_i$ in electrostatically confined QPCs strongly depend on the applied bias voltage in a non-universal way, probably due to charging effects. Therefore one would like to exclude this dependence from the quantities used for comparison with experiment.

Consider the ratio of the excess noise to the tunneling rate:
\eq{
\label{noise_tun}X = \frac{\tilde{S}(0)}{r} = e I_s \frac{\sum_{i = 1}^{3} \theta_i F_i}{\sum_{i = 1}^{3} \theta_i G_i} = e I_s \frac{F_1 + \theta F_3}{G_1 + \theta G_3},
}
where $\theta_i = |\eta_i|^2 (v_c/v_n)^{2 ((\mathbf{g}_i)_2)^2}$, $\theta = \theta_{3}/(\theta_{1} + \theta_{2})$, $v_c$ and $v_n$ are the propagation velocities of the charged and the neutral mode respectively, and $e$ is the elementary charge. The number $(\mathbf{g}_i)_2$ is presented in the column $g_2$ of Table~\ref{tab:exc} for each of the three excitations enumerated by $i$. Functions $F_i$ and $G_i$ (see Appendices B1, C1, C2 and E) represent contribution of different excitations to the excess noise and tunneling current respectively. In particular, the excess noise is given by
\eq{\label{noise_expr}\tilde{S}(0) = \frac{4 e^2 (\pi T_s)^{4\delta - 1}}{\hbar^{4 \delta + 1} v_c^{4 \delta}} \sum_i \theta_i F_i,}
and the tunneling rate is equal to
\eq{\label{tunn_coeff_expr}r = \frac{4 e (\pi T_s)^{4\delta - 1}}{I_s \hbar^{4 \delta + 1} v_c^{4 \delta}} \sum_i \theta_i G_i.}

The explicit form of these functions is presented below. Note that $F_1 = F_2$ and $G_1 = G_2$.

\eq{
\label{G_i}
G_i = \sin{2\pi\delta} \int\limits_{0}^{\infty}dt \frac {Q_i \lambda^{2 \delta}\,\sin Q_i j_s t}{ (\sinh t)^{2\delta}\, (\sinh\lambda t)^{2\delta}}
}
\eq{
\label{F_i}
F_i = F^{TT}_{i} \cos{2\pi\delta} - \frac{2}{\pi} F^{0T}_i \sin{2\pi\delta},
}
\nmq{
\label{F_TT}
F^{TT}_{i} = Q_i^2 \times\\ \times \lim_{\varepsilon \rightarrow +0} \left(\frac{\varepsilon^{1-4\delta}}{1-4\delta} + \int\limits_{\varepsilon}^{\infty}dt \frac {\lambda^{2\delta}\,\cos Q_i j_s t}{ (\sinh t)^{2\delta}\, (\sinh\lambda t)^{2\delta}} \right)
}
\eq{
\label{F_0T}
F^{0T}_{i} = \int\limits_{0}^{\infty}dt \frac {Q_i^2\lambda^{2\delta}\, t\cos Q_i j_s t}{ (\sinh t)^{2\delta}\, (\sinh\lambda t)^{2\delta}}
}
\eq{
\label{j_s}
j_s = \frac{I_s}{I_0},\quad I_0 = \nu \frac{e}{h} \pi k_B T_s = \nu \frac{e}{h} \pi k_B T_0,
}
where $\lambda = T_n/T_s$, $T_n$ is the local upper edge temperature at \textit{QPC}, $T_s = T_0$ is the local lower edge temperature at \textit{QPC}, $e$ is the elementary charge, $h = 2 \pi \hbar$ is the Planck constant, $k_B$ is the Boltzmann constant, $\nu = 2/3$ is the filling factor, and the scaling dimension $\delta$ and the quasiparticle charges in the units of the elementary charge $Q_i$ can be found in Table~\ref{tab:exc}.

\subsubsection{Remarks on non-universality in the noise to tunneling rate ratio}

It is easy to see that if any one quasiparticle dominates tunneling (for example, if $\theta \rightarrow \infty$) then the unwanted non-universal dependence of the tunneling amplitudes on the applied bias voltage does not enter the expression \eqref{noise_tun}. If we assume that the SU(2) symmetry of the edge \cite{KaneFisher, KaneFisherPolchinski} is for some reason preserved at the tunneling contact so that $|\eta_1|^2 = |\eta_2|^2 = |\eta_3|^2$, then again the non-universal behavior of the tunneling amplitudes does not enter the expression $X$; moreover, in this case finding $\theta$ allows us to determine the $v_c/v_n$ ratio. In general, though, $\theta$ may exhibit some non-universal behavior. Anticipating results, we can say that, surprisingly, $\theta$ does not seem to exhibit any strong dependence on $I_s$ or $I_n$.

For the following considerations we also give the large-$I_s$ asymptotic behavior of the noise to tunneling rate ratio \eqref{noise_tun} which we derive using Eqs.~\eqref{G_i}-\eqref{j_s}:
\nmq{\label{noise_tun_asymp}
\left.X_{\lambda, \theta}(I_s)\right|_{I_s \rightarrow \infty} = Q_1 e |I_s| \frac{1 + \theta(I_n, I_s) (Q_3/Q_1)^{4 \delta + 1}}{1 + \theta(I_n, I_s) (Q_3/Q_1)^{4 \delta}}\\
= \frac{e}{3} |I_s| \frac{1 + 2^{7/3}\theta(I_n, I_s)}{1 + 2^{4/3}\theta(I_n, I_s)}.
}
This asymptotic expression can give the reader an idea as to the effect introduced by the non-universal function $\theta(I_n, I_s)$. One can see, for example, that the gradient of the asymptote increases by a factor of 2 as $\theta$ increases from zero to infinity.

\section{Comparison with the experiment}

In this section we compare our analytical results with the experimental data.

The following data are available from the paper \cite{HeiblumExp}: the transmission rate $t = 1 - r$ dependence on the currents $I_n$ and $I_s$ (Fig.~3(a) of \cite{HeiblumExp}), the dependence of the excess noise at zero frequency on the currents $I_n$ and $I_s$ (Fig.~3(a) of \cite{HeiblumExp}) and the dependence of the excess noise at zero frequency on the current $I_n$ for $I_s = 0$ (Fig.~2 of \cite{HeiblumExp}).

It is a well-known problem (see, e.g., \cite{TunnellingRate0, TunnellingRate1, TunnellingRate2} and references therein) that the dependence of the transmission rate $t$ on the current $I_s$ does not have the form predicted by the minimal model of tunneling defined in Eqs.~\eqref{tun_ham_rel}, \eqref{tun_oppsA}. A possible explanation is the non-universal dependence of the tunneling amplitudes $\eta_i$ on $I_s$ due to electrostatic effects. As discussed in the previous section this problem can be avoided in simple cases by considering the ratio of the excess noise to the tunneling rate $r = 1 - t$. However, in the present case a certain degree of non-universality remains due to the non-universal function $\theta(I_n, I_s)$. The theoretical expression for the noise to tunneling rate ratio $X_{\lambda, \theta}(I_s)$ is given by Eq.~\eqref{noise_tun}, where $\lambda = T_n/T_s$ is the ratio of the two edges' temperatures. Neither $\theta$, nor $\lambda$ can be calculated theoretically and we will deduce them from fits of the experimental data. We assume that $\lambda$ depends on $I_n$ but not on $I_s$; we also assume that the non-universal behavior of the tunneling amplitudes does not lead to any significant dependence of $\theta$ on the currents $I_n, I_s$. While the former assumption is physically plausible in the weak tunneling regime, the latter one is motivated by our intention to reduce the number of fitting parameters as much as we can.

\begin{figure}[h!]
  \center{\includegraphics[width=1\linewidth]{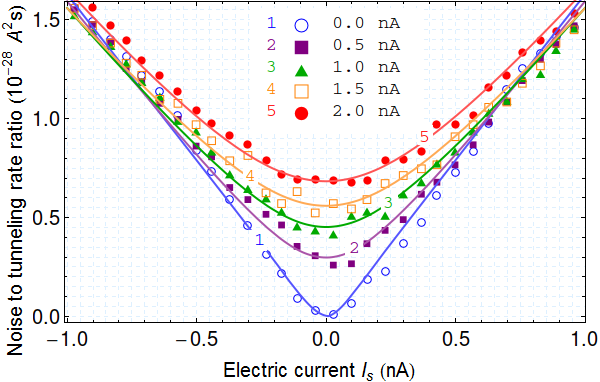}}
  \caption{\textbf{Excess noise to tunneling rate ratio as a function of the current $I_s$.} (Color online). Shown are experimental points and fits thereof by theoretical curves for different values of the current $I_n$. The legend shows the $I_n$ value in nA for each curve (plot symbol). Fitting parameters $\lambda$ and $\theta$ are defined independently for each value of $I_n$.}
  \label{fig:fit1}
\end{figure}

In Fig.~\ref{fig:fit1} the results of fitting $X_{\lambda, \theta}(I_s)$ to the experimental data taken from Fig.~3(a) of \cite{HeiblumExp} are shown. Optimal fits are found for each set of points corresponding to a given value of $I_n$ with $\theta$ and $\lambda$ being fitting parameters. The corresponding values and standard deviations of fitting parameters are shown in Table~\ref{thetalambdaTable}. For $I_n = 0$ we have set $\lambda = 1$ by definition.

\begin{table}[h!]
  \caption{Results of fitting the experimental points from Fig.~3(a) of \cite{HeiblumExp} by the function $X_{\lambda, \theta}(I_s)$ defined in Eq.~\eqref{noise_tun}. Fitting parameters $\lambda$ and $\theta$ are defined independently for each value of current $I_n$. $\Delta\lambda$ and $\Delta\theta$ are standard deviations of $\lambda$ and $\theta$ respectively.}
  \label{thetalambdaTable}
  \begin{center}
  \begin{tabular}{|c|c|c|c|c|c|}
  \hline
  \# & $ \quad I_n\,\mathrm{(nA)} \quad$ & $\qquad \lambda \qquad$ & $\quad \Delta\lambda \quad$ & $\quad\theta\quad$ & $\quad\Delta\theta\quad$ \\
  \hline
  1 &  $0.0$& $1.00$& $-$ & $0.53$ & $0.04$\\
  \hline
  2 &  $0.5$& $4.48$& $0.19$& $0.44$ & $0.03$\\
  \hline
  3 &  $1.0$& $6.16$& $0.15$& $0.35$ & $0.02$\\
  \hline
  4 &  $1.5$& $7.32$& $0.17$& $0.30$ & $0.02$\\
  \hline
  5 &  $2.0$& $8.65$& $0.13$& $0.36$ & $0.03$\\
  \hline
  \end{tabular}
  \end{center}
\end{table}

\begin{figure}[h!]
  \center{\includegraphics[width=1\linewidth]{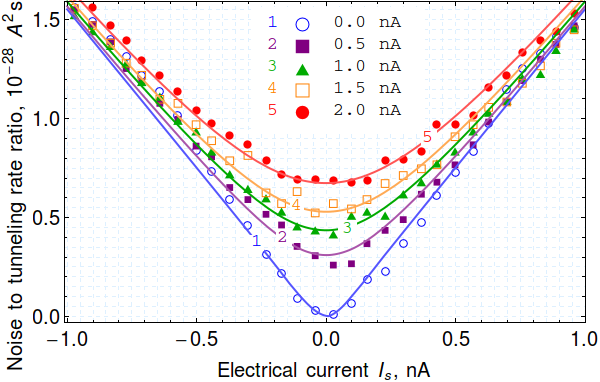}}
  \caption{\textbf{Excess noise to tunneling rate ratio as a function of the current $I_s$.} (Color online). Shown are experimental points and fits thereof by theoretical curves for different values of the current $I_n$. The legend shows the $I_n$ value in nA for each curve (plot symbol). Fitting parameter $\lambda$ is defined independently for each value of $I_n$. Parameter $\theta$ is set to $\theta = \theta_{\mathrm{mean}} = \sum_i\theta_i/5 = 0.39$.}
  \label{fig:fit2}
\end{figure}

As one can see from the Table~\ref{thetalambdaTable}, the values of $\theta$ do not vary significantly. Thus we repeat the fitting procedure with $\theta$ equal to the mean of the five values and $\lambda$ being the only fitting parameter. The resulting fits and values of $\lambda$ are presented in Fig.~\ref{fig:fit2} and Table~\ref{onlylambdaTable}. As one can see the fits remain good, thus we cannot reliably find the extent of deviation of $\theta$ from a constant value with the available experimental data.

\begin{table}[h!]
  \caption{Results of fitting the experimental points from Fig.~3(a) of \cite{HeiblumExp} by the function $X_{\lambda, \theta}(I_s)$ defined in Eq.~\eqref{noise_tun}. Fitting parameter $\lambda$ is defined independently for each value of current $I_n$ for fixed $\theta = \theta_{\mathrm{mean}} = \sum_i\theta_i/5 = 0.39$. $\Delta\lambda$ is the standard deviation of $\lambda$.}
  \label{onlylambdaTable}
  \begin{center}
  \begin{tabular}{|c|c|c|c|}
  \hline
  \# & $ \quad I_n\,\mathrm{(nA)} \quad$ & $\qquad \lambda \qquad$ & $\quad \Delta\lambda \quad$\\
  \hline
  1 &  $0.0$& $1.00$& $-$\\
  \hline
  2 &  $0.5$& $4.62$& $0.18$\\
  \hline
  3 &  $1.0$& $5.98$& $0.14$\\
  \hline
  4 &  $1.5$& $6.99$& $0.17$\\
  \hline
  5 &  $2.0$& $8.55$& $0.11$\\
  \hline
  \end{tabular}
  \end{center}
\end{table}

\begin{figure}[h!]
  \center{\includegraphics[width=1\linewidth]{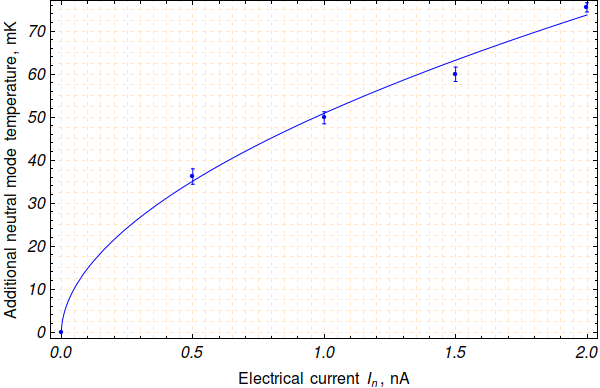}}
  \caption{\textbf{Excess temperature of the upper edge $T_n - T_0$ as a function of the current $I_n$.} (Color online). Comparison of the points obtained from the data in Table~\ref{onlylambdaTable} with the fit of these points by formula \eqref{temp_model} is shown.}
  \label{fig:TempCurr}
\end{figure}

Table~\ref{onlylambdaTable} gives us some data on the dependence of $T_n = \lambda T_s = \lambda T_0$ on the current $I_n$. We further investigate this by fitting it with the following function:
\eq{\label{temp_model}T_n = T_0\Biggl(1 + C\ |I_n|^a\Biggl),}
where $C$ and $a$ are fitting parameters. The resulting fit is shown in Fig.~\ref{fig:TempCurr}. The corresponding values of fitting parameters are $a = 0.54(5)$, $C = 5.05(13)\,\mathrm{nA}^{-a}$. This disagrees with the claim of Ref.~\cite{Rosenow} that the experimental data are consistent with a linear $T_n$ dependence on $I_n$. We cannot analyze the source of this discrepancy because Ref.~\cite{Rosenow} does not contain sufficient detail as to how the comparison with the experiment was done.

\begin{figure}[h!]
  \center{\includegraphics[width=1\linewidth]{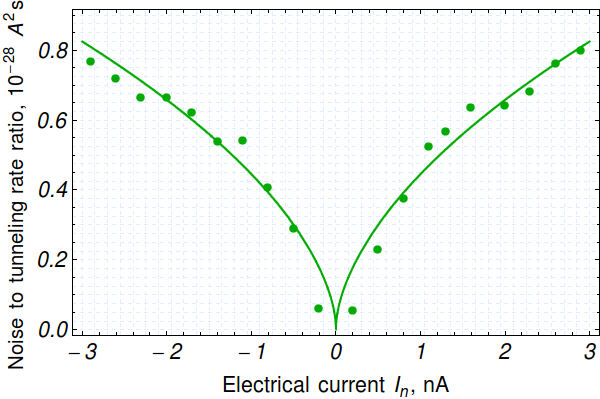}}
  \caption{\textbf{Excess noise to tunneling rate ratio for $I_s = 0$ as a function of current $I_n$.} (Color online). Experimental points are taken from Fig.~2 of \cite{HeiblumExp} for the tunneling rate $r \approx 0.2$. The theoretical curve is obtained for $\theta = \theta_{mean} = \sum_i\theta_i/5 = 0.39$. The values of $\lambda$ are given by Eq.~\eqref{temp_model}. No fitting procedure is involved.}
  \label{fig:NoiseDepOnIn}
\end{figure}

Using the phenomenological dependence \eqref{temp_model} it is possible to predict the noise to tunneling rate ratio at any $I_s$, $I_n$ without any further fitting procedures (we still take $\theta = \theta_{\mathrm{mean}} = 0.39$). So we can test the formula \eqref{temp_model} by comparing the theoretical prediction of $X_{\lambda, \theta}(I_s)$ to another data set. We take the experimental data for the excess noise $\tilde{S}(0)$ dependence on $I_n$ for $I_s = 0$ from Fig.~2 of the paper \cite{HeiblumExp} for $t = 1-r = 80\%$. The resulting comparison of the noise to tunneling rate ratio $X$ is shown in Fig.~\ref{fig:NoiseDepOnIn}. An excellent agreement of theoretical curves and experimental points gives an independent confirmation of the result \eqref{temp_model}.

\section{Discussion}

In this section we discuss the results of the comparison of theoretical predictions with the experimental data, and emphasize some important aspects of our analysis.

The good quality of the fits shown in Fig.~\ref{fig:fit2} suggests that the minimal model of the $\nu = 2/3$ quantum Hall edge is consistent with the experimental data. Note, that the existence of good fits is not trivial because of the following reasons. The number of fitting parameters is small; namely, two fitting parameters are used to get Fig.~\ref{fig:fit1}, only one is used for Fig.~\ref{fig:fit2} and no fitting parameters are involved in obtaining Fig.~\ref{fig:NoiseDepOnIn}. Moreover, our theory imposes strong constraints on the shape of the function $X_{\lambda, \theta}(I_s)$ in the whole region of parameters $\lambda$, $\theta$. For example, as can be seen from Eq.~\eqref{noise_tun_asymp}, the gradient of the large $I_s$ asymptote of the curve $X_{\lambda, \theta}(I_s)$ varies between $e/3$ and $2e/3$ as $\theta$ increases from zero to infinity. The fact that the experimental curve lies between these bounds is non-trivial.

The fact that the gradient of the large $I_s$ asymptote of the curve $X_{\lambda, \theta}(I_s)$ does not coincide with the limiting values of $e/3$ and $2e/3$ provides an indirect confirmation of the presence of more than one quasiparticle species taking part in tunneling. Indeed, in the case of a single quasiparticle species of charge $Q$, the asymptote gradient would equal $Q$. From considerations similar to the flux insertion argument \cite[Section 7.5]{QHE_Book} one can deduce that the natural $\nu = 2/3$ fractional charges are integer multiples of $e/3$. Interpreting the asymptotic behavior of the noise to tunneling rate ratio in terms of a single quasiparticle tunneling, one would get an unnatural value of $Q$ lying between $e/3$ and $2e/3$. The minimal model of the $\nu = 2/3$ edge explains this contradiction in a natural way: the experimentally observed "charge" is a weight average of the charges of two equally relevant quasiparticles with weights defined by non-universal tunneling amplitudes' ratio $\theta$. This gives an extra argument in favor of the minimal model of the $\nu = 2/3$ edge with the K-matrix \eqref{K_min}. A similar point was made in paper \cite{Ferraro} in relation to the experiment \cite{HeiblumForFerraro}.

Note, that the minimal model analyzed here can also be regarded as the low-energy limit of the extended models proposed in Refs.~\cite{Ferraro, MeirGefen23EdgeReconstruction}. At higher energies both extended models predict tunneling contact physics to be dominated by a quasiparticle with charge $e/3$. Since we do not see this in our analysis, we conclude that either the extended physics is not present in the system or occurs above the energies probed in the experiment of Ref.~\cite{HeiblumExp}.

It should, however, be emphasized that the minimal $\nu = 2/3$ edge model alone is not sufficient to describe the present experiment. Extra assumptions are needed to model the non-universal physics of Ohmic contacts, edge equilibration mechanisms, and the tunneling contact. Such assumptions have been discussed throughout the text, and here we summarize them:
\begin{itemize}
  \item injection of electric current into an Ohmic contact induces non-equilibrium noise in the neutral mode but not in the charged mode;
  \item injection of electric current into an Ohmic contact does not induce a shift in the neutral mode chemical potential (that is the thermodynamic potential dual to the neutral charge defined through Eqs.~\eqref{neut_current operator} and \eqref{neut_charge_vector});
  \item strong equilibration of the charged and the neutral modes takes place along the edge resulting in some current-dependent local temperature of the edge;
  \item the tunneling contanct can be modeled by the minimal tunneling Hamiltonian \eqref{tun_ham_rel} with tunneling amplitudes depending on the edge chemical potential in some non-universal way.
\end{itemize}
While these phenomenological assumptions are plausible, they may not be accurate. Moreover, their validity may depend on the experimental conditions.

The theoretical framework presented here enables a more detailed experimental investigation and refinement of our understanding of non-equilibrium processes at the edge. For example, in the present work we use experimental data to establish a phenomenological law \eqref{temp_model} describing the dependence of the neutral mode temperature at the QPC on the current $I_n$ (see Figs.~\ref{fig:TempCurr} and Fig.~\ref{fig:NoiseDepOnIn}). Recently there has been some theoretical progress in understanding of the interaction of Ohmic contacts with the quantum Hall edge \cite{OhmicContactsLevkivsk}. However, at present a complete theoretical predictive model of Ohmic contacts is still missing, and the information on the neutral mode heating may contribute to its development.

It is also interesting to note that we do not find any significant dependence of the ratio of the tunneling amplitudes of different species of quasiparticles on the currents $I_n$, $I_s$ (see discussion of Figs.~\ref{fig:fit1} and \ref{fig:fit2}). This is surprising since the tunneling amplitudes themselves appear to vary significantly to explain the tunneling rate dependence on $I_s$ observed in \cite{HeiblumExp}. This fact suggests the existence of a mechanism which ensures roughly equal participation of all three quasiparticles species in the tunneling. It is known \cite{KaneFisher, KaneFisherPolchinski} that disorder scattering at the edge enforces the SU(2) symmetry between the quasiparticle species. A similar mechanism might be responsible for the discussed phenomenon.

We emphasize that our theoretical predictions are derived in the limit of perturbatively weak tunneling of the quasiparticles. Therefore, the tunneling rate at which the comparison with the experimental data is made should be small enough so that our theory remains valid, but large enough in order to minimize statistical errors of the noise to tunneling rate ratio.

\section{Conclusions}

Using the chiral Luttinger liquid theory of the quantum Hall edge we develop a quantitative model of the experiment reported in \cite{HeiblumExp}. This model enables us to extract important quantitative information about non-equilibrium processes in Ohmic and tunneling contacts from the experimental data. In particular, for $\nu = 2/3$, we find a power-law dependence of the neutral mode temperature on the charge current injected from the Ohmic contact. We also find a surprising behavior of quasiparticle tunneling amplitudes which may be a signature of the SU(2) symmetry in the quasiparticle tunneling across the QPC.

\section{Acknowledgements}

The research leading to these results has received funding from the European Research Council under the European Union's Seventh Framework Programme (FP7/2007-2013) / ERC grant agreement No 279738 - NEDFOQ.

\appendix

\section{Appendix A: Useful one-edge correlation functions}

Here we give explicit expressions for the correlation functions at a single edge without tunneling (described by the minimal model for $\nu = 2/3$ defined in the main text) which are used to calculate the quantities of experimental interest. In all the correlation functions of this appendix we assume the infinite system size limit $L \rightarrow \infty$.

The two-point correlation function of quasi-particle operators is equal to
\nmq{\label{2QP_CorrFunc_1}
\bigl\langle V_{\mathbf{g}}^{\dag} (x_1,t_1)  V_{\mathbf{g}'}(x_2,t_2)\bigl\rangle = \bigl\langle V_{-\mathbf{g}} (x_1,t_1)  V_{\mathbf{g}'}(x_2,t_2)\bigl\rangle =\\
 = \delta_{\mathbf{g},\mathbf{g}'}\prod_{p = c, n} F_p(x_1-x_2, t_1-t_2 - i \varepsilon,\mathbf{g})
}
\eq{\label{2QP_CorrFunc_2}
F_p(x, t, \mathbf{g}) = \frac{(\pi T)^{g_p^2}}{(i v_{p} \sinh{\pi T X_p})^{g_p^2}} \exp{(i Q^{(p)} \mu^{(p)} X_p/v_p)}
}
where $V_{\mathbf{g}}(x,t)$ is a quasiparticle excitation operator defined in Eq.~\eqref{vertex_quasi}, $\mathbf{g} = (g_1, g_2) = (g_c, g_n)$ is the excitation vector, $p$ enumerates charged ($c$ or $1$) and neutral ($n$ or $2$) modes, $X_p = -\chi_p x+v_p t$, $\chi_p$ and $v_p$ are the mode chirality and velocity respectively which enter the action \eqref{action} (in our case $\chi_1 = -\chi_2 = 1$), and $T$ is the temperature of the edge. The electric charge $Q^{(c)} = Q = g_1' \sqrt{\nu}$, $\nu = 2/3$, and $\mu^{(c)} = \mu$ is the chemical potential of the charged mode at the edge. It coincides with the chemical potential of the Ohmic contact where the charged mode originates. The neutral charge $Q^{(n)} = g_2'$ and the chemical potential $\mu^{(n)}$ do not enter the formulas in the other sections as we assume $\mu^{(n)} = 0$, though, in principle, injection of the current from an Ohmic contact could shift the neutral mode chemical potential. We have also introduced an infinitesimally small positive number $\varepsilon \rightarrow +0$.

The electric current along the edge in equilibrium is given by the average of the current operator $J^{\mu = 1}$ defined in Eq.~\eqref{el_current operator_2}:
\nmq{\left\langle J^{1}(x,t) \right\rangle = \chi_c v_c \left\langle J^{0}(x,t) \right\rangle =\\
 = - v_c \frac{\sqrt{\nu}}{L} \left\langle \pi^0_{(c)} \right\rangle = \chi_c \frac{\nu}{2 \pi} \mu^{(c)} = \frac{\nu}{2 \pi} \mu^{(c)}}
in agreement with the quantization law of Hall conductance \cite{CheianovFrolichAlekseev}.

The two-point correlation function of quasi-particle operators with the current operator inserted is given by
\nmq{\label{2QP+Curr_CorrFunc_1}
\bigl\langle J^{1}(x_0,t_0) V_{\mathbf{g}}^{\dag} (x_1,t_1)  V_{\mathbf{g}'}(x_2,t_2)\bigl\rangle =\\
 = \bigl\langle V_{\mathbf{g}}^{\dag} (x_1,t_1)  V_{\mathbf{g}'}(x_2,t_2)\bigl\rangle \times \Bigl( \bigl\langle J^{1}(x_0,t_0) \bigl\rangle +\\
  + \frac{Q^{(c)} \chi_c \pi T}{2\pi i}\left(\coth{\pi T(Y_0-Y_1)} - \coth{\pi T(Y_0-Y_2)}\right) \Bigl),
}
where $Y_i = t - \chi_c x/v_c + i \kappa_i$. $\kappa_0 = 0$, $\kappa_1 = \kappa \rightarrow +0$ is an infinitesimally small positive number, $\kappa_2 = \kappa_1 + \varepsilon$, and $\varepsilon$ is the same as in the two-particle correlation function.

Finally, the current-current correlation function is
\nmq{\label{2Curr_CorrFunc_1}
\bigl\langle J^{1}(x_0,t_0) J^{1}(x_1,t_1)\bigl\rangle = \bigl\langle J^{1}(x_0,t_0) \bigl\rangle \bigl\langle J^{1}(x_1,t_1)\bigl\rangle +\\
 + \frac{\nu}{(2\pi)^2} \frac{(\pi T)^2}{(i \sinh{\pi T (Y_0 - Y_1)})^2},
}
where $Y_i = t - \chi_c x/v_c + i \kappa_i$, $\kappa_0 = 0$, $\kappa_1 = \kappa \rightarrow +0$ is an infinitesimally small positive number.

\section{Appendix B: Tunneling current}

Here we present a derivation of the expressions for the tunneling current $I_{T}$ and the tunneling rate $r$.

The tunneling current can be defined as the time derivative of the total charge at the lower edge:
\begin{eqnarray}
I_{T} = \frac {d}{dt}Q^{(l)} &=& i [H, Q^{(l)}] = i [H_T , Q^{(l)}],\\
Q^{(l)} &=& \int\limits_{-\infty}^{\infty} J^{0 (l)}(x,t) dx.
\end{eqnarray}
Here $J^{0 (l)}$ is the lower edge charge density operator $J^0$ defined in Eq.~\eqref{el_current operator_2}, $H$ is the full system Hamiltonian and $H_T$ is the tunneling Hamiltonian \eqref{tun_ham_rel}. Using the latter we get an explicit expression
\eq{I_{T, \mathrm{int}}(t) = i \sum_i Q_i\bigl(\eta_i A_i(t) - \eta_i^* A_i^\dag(t) \bigl),}
where $Q_i$ are the quasiparticle charges $Q$ in Table~\ref{tab:exc} and $A_i$ are the operators defined in Eq.~\eqref{tun_oppsA}. This is the tunneling current operator in the \textit{interaction picture} with interaction $H_T$ (which is emphasized by the subscript "$\mathrm{int}$"). We calculate the expression for the tunneling current operator in the \textit{Heisenberg picture} within the perturbation theory in $H_T$:
\nmq{
\label{tun_curr} I_{T}(t) = I_{T, \mathrm{int}}(t)+i\int\limits_{-\infty}^t d\tau \left[H_T(\tau), I_{T, \mathrm{int}}(t)\right] + O(|\eta_i|^3) =\\
 = i \sum_i Q_i\bigl(\eta_i A_i(t) - \eta_i^* A_i^\dag(t) \bigl) -\\
  - \sum_{i, j} Q_i \int\limits_{-\infty}^t d\tau \left[\eta_j A_j(\tau) + \eta_j^* A_j^\dag(\tau), \eta_i A_i(t) - \eta_i^* A_i^\dag(t) \right] +\\
   + O(|\eta_i|^3).
}

The observed tunneling current is then
\nmq{\label{tun_curr_2_sum}\left\langle I_{T}(t) \right\rangle =\\
 = \sum_{i} Q_i |\eta_i|^2 \int\limits_{-\infty}^t d\tau \left\langle \left[A_i(\tau), A_i^{\dag}(t)\right] - \left[A_i^{\dag}(\tau), A_i(t)\right] \right\rangle +\\
  + O(|\eta_i|^3).}
We have used the relationships $\left\langle A_i(t) \right\rangle = \left\langle A_j(\tau) A_i(t) \right\rangle = 0$, $\left\langle A_j^{\dag}(\tau) A_i(t) \right\rangle \propto \delta_{ij}$.

It can be checked with explicit correlation functions \eqref{2QP_CorrFunc_2} that the integral of each of the summands in the formula \eqref{tun_curr_2_sum} is convergent. Thus, one can split them and manipulate separately. Using time translational invariance of the correlation functions in both summands and changing sign of the integration variable in the second one we finally get
\eq{
\label{tun_curr_expect_gen}
\left\langle I_{T}(t) \right\rangle = \sum_{i} Q_i |\eta_i|^2 \int\limits_{-\infty}^{+\infty} d\tau \left\langle \left[A_i(\tau), A_i^{\dag}(0)\right]\right\rangle + O(|\eta_i|^3),
}
which leads to the expression \eqref{tun_general} for the tunneling rate~$r$.

\section{Appendix B1: Tunneling current (continued)}

Starting from the expression \eqref{tun_curr_expect_gen} for the tunneling current expectation value and using the explicit form of the correlation functions \eqref{2QP_CorrFunc_1} and \eqref{2QP_CorrFunc_2}, we obtain up to corrections of $O(|\eta_i|^3)$
\begin{widetext}
\eq{
\left\langle I_{T}(t) \right\rangle = 2 i \sum_{i} Q_i |\eta_i|^2 v_c^{- 4 \delta} \left(\frac{v_c}{v_n}\right)^{2 ((\mathbf{g}_i)_2)^2} \int\limits_{-\infty}^{+\infty} d\tau \frac{(\pi T_n)^{2 \delta} (\pi T_s)^{2 \delta} \sin{Q_i \Delta\mu \tau}}{(i \sinh{\pi T_n (\tau - i \varepsilon)})^{2 \delta} (i \sinh{\pi T_s (\tau - i \varepsilon)})^{2 \delta}},
}
\end{widetext}
where $T_n = T^{(u)}$ is the upper edge temperature, $T_s = T^{(l)}$ is the lower edge temperature, $\Delta\mu = \mu^{(c, u)} - \mu^{(c, l)}$ is the difference of the chemical potentials of the upper and the lower edges' charged modes, the numbers $(\mathbf{g}_i)_1$, $(\mathbf{g}_i)_2$ are presented in the columns $g_1$, $g_2$ respectively of Table~\ref{tab:exc} for each of the three excitations enumerated by $i$, and $\delta$ is the scaling dimension of the excitations presented in the column $\delta$ of Table~\ref{tab:exc}, and $\varepsilon \rightarrow +0$ is an infinitesimally small positive number.

For $0 < \delta < 1/2$ the last formula can be further simplified:
\begin{widetext}
\eq{
\label{I_T_answer}
I_{T} = \left\langle I_{T}(t) \right\rangle = \sum_{i} 4 Q_i |\eta_i|^2 v_c^{- 4 \delta} \left(\frac{v_c}{v_n}\right)^{2 ((\mathbf{g}_i)_2)^2} \sin{2 \pi \delta} \int\limits_{0}^{+\infty} d\tau \frac{(\pi T_n)^{2 \delta} (\pi T_s)^{2 \delta} \sin{Q_i \Delta\mu \tau}}{(\sinh{\pi T_n \tau})^{2 \delta} (\sinh{\pi T_s \tau})^{2 \delta}}.
}
\end{widetext}

\section{Appendix C: Noise}

In this section we derive expressions for the noise spectral density $S(\omega)$ at zero frequency $\omega$.

The operator $I(t)$ of the full current flowing to the \textit{Voltage probe} can be presented as a sum of the tunneling current $I_T(t)$ defined in Eq.~\eqref{tun_curr} and the current $I_0$ flowing along the lower edge just before the \textit{QPC}:
\begin{eqnarray}
I(t) &=& I_0(t) + I_T(t),\\
I_0(t) &=& J^{1 (l)}(x = -0, t),
\end{eqnarray}
here $I(t)$ and $I_0(t)$ are operators in the \textit{Heisenberg picture}.

The noise spectral density $S(\omega)$ defined in Eq.~\eqref{noise_general} then separates into four terms, see Eqs.~\eqref{Noise_sep_1} and \eqref{Noise_sep_2}, where the identity $S_{ab}(\omega) = S_{ba}(-\omega)$ following from the time translational invariance of the correlation functions has been used.

Using Eq.~\eqref{2Curr_CorrFunc_1} one obtains
\nmq{
S_{00}(\omega = 0) =\\
 = \frac 12 \frac{\nu}{(2\pi)^2} \int\limits_{-\infty}^\infty d\tau \frac{(\pi T^{(l)})^2}{(i \sinh{\pi T^{(l)} (-\tau - i \varepsilon)})^2} + \mathrm{c.c.} =\\
  = \frac{\nu}{2\pi} T^{(l)},
}
where $T^{(l)}$ is the lower edge temperature, and $\varepsilon \rightarrow +0$ is an infinitesimally small positive number. This is the identity \eqref{noise_00}.

Since $\left\langle A_i(t) \right\rangle = 0$, $S_{TT}(\omega)$ can be expressed in the following way up to corrections $O(|\eta_i|^3)$:
\eq{S_{TT}(\omega) = \int\limits_{-\infty}^\infty d\tau \exp\bigl(i\omega\tau \bigl) \frac 12 \Bigl\<\Bigl\{I_{T}(0), I_{T}(\tau)\Bigl\}\Bigl\>.}
Using $\left\langle A_j(\tau) A_i(t) \right\rangle = 0$, $\left\langle A_j^{\dag}(\tau) A_i(t) \right\rangle \propto \delta_{ij}$ and neglecting terms $O(|\eta_i|^3)$ we further simplify this expression to
\nmq{
S_{TT}(\omega) =\\
 = \frac 12 \sum_i Q_i^2 |\eta_i|^2 \int\limits_{-\infty}^\infty d\tau \exp\bigl(i\omega\tau \bigl) \left\langle \left\{A_i(0), A_i^\dag(\tau)\right\} \right\rangle +\\
  + \mathrm{c.c.},
}
which at $\omega = 0$ is equivalent to Eq.~\eqref{noise_TT} due to the time translational invariance of the correlation functions.

Moving to $S_{0T}(\omega)$, we find up to the corrections $O(|\eta_i|^3)$ that
\nmq{
S_{0T}(\omega) = \frac 12 \sum_i Q_i |\eta_i|^2 \int\limits_{-\infty}^\infty d\tau \int\limits_{-\infty}^{\tau} d\tau' \exp\bigl(i\omega\tau \bigl)\times\\
 \times \left\langle\left\{\Delta I_{0}(0), \left[A_i(\tau'), A_i^{\dag}(\tau)\right] - \left[A_i^{\dag}(\tau'), A_i(\tau)\right]\right\}\right\rangle.
}
In analogy with the calculation of the tunneling current expectation value, the integral of each of the two summands in the last formula is convergent, thus we can manipulate the two summand integrals separately. Changing the order of integration and renaming $\tau \leftrightarrow \tau'$ in the second summand we arrive at the expression \eqref{noise_0T} for $\omega = 0$.

\section{Appendix C1: Noise~--- the TT term}

Starting from the expression \eqref{noise_TT} for the $TT$ component of the current noise and using the explicit form of the correlation functions \eqref{2QP_CorrFunc_1} and \eqref{2QP_CorrFunc_2}, we obtain up to corrections of $O(|\eta_i|^3)$
\begin{widetext}
\eq{
S_{TT}(0) = 2 \sum_{i} Q_i^2 |\eta_i|^2 v_c^{- 4 \delta} \left(\frac{v_c}{v_n}\right)^{2 ((\mathbf{g}_i)_2)^2} \int\limits_{-\infty}^{+\infty} d\tau \frac{(\pi T_n)^{2 \delta} (\pi T_s)^{2 \delta} \cos{Q_i \Delta\mu \tau}}{(i \sinh{\pi T_n (\tau - i \varepsilon)})^{2 \delta} (i \sinh{\pi T_s (\tau - i \varepsilon)})^{2 \delta}},
}
\end{widetext}
where $T_n = T^{(u)}$ is the upper edge temperature, $T_s = T^{(l)}$ is the lower edge temperature, $\Delta\mu = \mu^{(c, u)} - \mu^{(c, l)}$ is the difference of the chemical potentials of the upper and the lower edges' charged modes, the numbers $(\mathbf{g}_i)_1$, $(\mathbf{g}_i)_2$ are presented in the columns $g_1$, $g_2$ respectively of Table~\ref{tab:exc} for each of the three excitations enumerated by $i$, and $\delta$ is the scaling dimension of the excitations presented in the column $\delta$ of Table~\ref{tab:exc}, and $\varepsilon \rightarrow +0$ is an infinitesimally small positive number.

For $0 < \delta < 3/4$ the last formula can be rewritten as
\begin{widetext}
\eq{
\label{S_TT_answer}
S_{TT}(0) = 4 \sum_{i} Q_i^2 |\eta_i|^2 v_c^{- 4 \delta} \left(\frac{v_c}{v_n}\right)^{2 ((\mathbf{g}_i)_2)^2} \cos{2 \pi \delta} \times \lim_{\varepsilon \rightarrow +0} \left(\int\limits_{\varepsilon}^{+\infty} d\tau \frac{(\pi T_n)^{2 \delta} (\pi T_s)^{2 \delta} \cos{Q_i \Delta\mu \tau}}{(\sinh{\pi T_n \tau})^{2 \delta} (\sinh{\pi T_s \tau})^{2 \delta}} + \frac{\varepsilon^{1-4\delta}}{1-4\delta}\right).
}
\end{widetext}

\section{Appendix C2: Noise~--- the 0T term}

Starting from the expression \eqref{noise_0T} for the $0T$ component of the current noise and using the explicit form of the correlation functions \eqref{2QP_CorrFunc_1}, \eqref{2QP_CorrFunc_2}, and \eqref{2QP+Curr_CorrFunc_1}, we obtain up to corrections of $O(|\eta_i|^3)$

\begin{widetext}
\nmq{
S_{0T}(0) = \sum_{i} \frac{Q_i^2}{2 \pi} |\eta_i|^2 v_c^{- 4 \delta} \left(\frac{v_c}{v_n}\right)^{2 ((\mathbf{g}_i)_2)^2} \int\limits_{-\infty}^{+\infty} dt \int\limits_{-\infty}^{+\infty} d\tau \frac{i(\pi T_n)^{2 \delta} (\pi T_s)^{2 \delta + 1} \cos{Q_i \Delta\mu (\tau - t)}}{(i \sinh{\pi T_n (\tau - t - i (\kappa - \varepsilon))})^{2 \delta} (i \sinh{\pi T_s (\tau - t - i (\kappa - \varepsilon))})^{2 \delta}}\\ \times (\coth{\pi T_s (-\tau - i \varepsilon)} - \coth{\pi T_s (-t - i \kappa)}) + \mathrm{c.c.},
}
\end{widetext}
where $T_n = T^{(u)}$ is the upper edge temperature, $T_s = T^{(l)}$ is the lower edge temperature, $\Delta\mu = \mu^{(c, u)} - \mu^{(c, l)}$ is the difference of the chemical potentials of the upper and the lower edges' charged modes, the numbers $(\mathbf{g}_i)_1$, $(\mathbf{g}_i)_2$ are presented in the columns $g_1$, $g_2$ respectively of Table~\ref{tab:exc} for each of the three excitations enumerated by $i$, and $\delta$ is the scaling dimension of the excitations presented in the column $\delta$ of Table~\ref{tab:exc}, and $\varepsilon \rightarrow +0$, $\kappa \rightarrow +0$ are infinitesimally small positive numbers such that $\kappa > \varepsilon$.

It is tempting to integrate each of the two hyperbolic cotangents separately, however, the integrals of  a signle cotangent diverge as $t$  and $\tau$ go to $\pm\infty$ with $t-\tau$ being finite. Yet, the integral of the difference of the two cotangents is absolutely convergent. After a change of variables $\tau = t + y$ we get:
\begin{widetext}
\nmq{
S_{0T}(0) = \sum_{i} \frac{Q_i^2}{2 \pi} |\eta_i|^2 v_c^{- 4 \delta} \left(\frac{v_c}{v_n}\right)^{2 ((\mathbf{g}_i)_2)^2} \int\limits_{-\infty}^{+\infty} dy \int\limits_{-\infty}^{+\infty} dt \frac{i(\pi T_n)^{2 \delta} (\pi T_s)^{2 \delta + 1} \cos{Q_i \Delta\mu y}}{(i \sinh{\pi T_n (y - i (\kappa - \varepsilon))})^{2 \delta} (i \sinh{\pi T_s (y - i (\kappa - \varepsilon))})^{2 \delta}}\\
 \times (\coth{\pi T_s (-t - y - i \varepsilon)} - \coth{\pi T_s (-t - i \kappa)}) + \mathrm{c.c.}
}

Since
\nmq{
\int\limits_{-\infty}^{+\infty} dt (\coth{\pi T_s (-t - y - i \varepsilon)} - \coth{\pi T_s (-t - i \kappa)}) =
\int\limits_{-\infty}^{+\infty} dt (\coth{\pi T_s (t - y - i \varepsilon)} - \coth{\pi T_s (t - i \kappa)}) = \\
\left. \frac{1}{\pi T_s} \ln{\frac{\sinh{\pi T_s (t - y - i \varepsilon)}}{\sinh{\pi T_s (t - i \kappa)}}}\right|_{-\infty}^{+\infty} = - 2 (y - i (\kappa - \varepsilon)),
}
we get
\nmq{
S_{0T}(0) = \sum_{i} \frac{2 Q_i^2}{\pi} |\eta_i|^2 v_c^{- 4 \delta} \left(\frac{v_c}{v_n}\right)^{2 ((\mathbf{g}_i)_2)^2} \int\limits_{-\infty}^{+\infty} dy \frac{-i(\pi T_n)^{2 \delta} (\pi T_s)^{2 \delta + 1} (y - i (\kappa - \varepsilon)) \cos{Q_i \Delta\mu y}}{(i \sinh{\pi T_n (y - i (\kappa - \varepsilon))})^{2 \delta} (i \sinh{\pi T_s (y - i (\kappa - \varepsilon))})^{2 \delta}} =\\
 \sum_{i} \frac{2 Q_i^2}{\pi} |\eta_i|^2 v_c^{- 4 \delta} \left(\frac{v_c}{v_n}\right)^{2 ((\mathbf{g}_i)_2)^2} \int\limits_{-\infty}^{+\infty} dy \frac{-i(\pi T_n)^{2 \delta} (\pi T_s)^{2 \delta + 1} y \cos{Q_i \Delta\mu y}}{(i \sinh{\pi T_n (y - i (\kappa - \varepsilon))})^{2 \delta} (i \sinh{\pi T_s (y - i (\kappa - \varepsilon))})^{2 \delta}}.
}

For $0 < \delta < 1/2$ the last formula can be rewritten as
\eq{
\label{S_0T_answer}
S_{0T}(0) = - 4 \sum_{i} \frac{Q_i^2}{\pi} |\eta_i|^2 v_c^{- 4 \delta} \left(\frac{v_c}{v_n}\right)^{2 ((\mathbf{g}_i)_2)^2} \sin{2 \pi \delta} \int\limits_{0}^{+\infty} d\tau \frac{(\pi T_n)^{2 \delta} (\pi T_s)^{2 \delta + 1} \tau \cos{Q_i \Delta\mu \tau}}{(\sinh{\pi T_n \tau})^{2 \delta} (\sinh{\pi T_s \tau})^{2 \delta}}.
}
\end{widetext}

\section{Appendix D: Excess noise}

In the equilibrium ($\Delta\mu = 0$ and $T_n = T_s = T_0$) one can represent the integrals in formulas \eqref{S_TT_answer} and \eqref{S_0T_answer} in terms of Euler gamma function which leads to
\nmq{
S_{TT}(0)|_{\mathrm{eq}} = 4 \sum_{i} Q_i^2 |\eta_i|^2 v_c^{- 4 \delta} \left(\frac{v_c}{v_n}\right)^{2 ((\mathbf{g}_i)_2)^2} \cos{2 \pi \delta} \times\\ (\pi T_0)^{4 \delta - 1} \frac{1}{2 \sqrt{\pi}} \Gamma{\left(\frac{1}{2}-2\delta\right)} \Gamma{(2\delta)},
}
\nmq{
S_{0T}(0)|_{\mathrm{eq}} = -\frac{4}{\pi} \sum_{i} Q_i^2 |\eta_i|^2 v_c^{- 4 \delta} \left(\frac{v_c}{v_n}\right)^{2 ((\mathbf{g}_i)_2)^2} \sin{2 \pi \delta} \times\\ (\pi T_0)^{4 \delta - 1} \frac{\sqrt{\pi}}{4} \cot{(2 \pi \delta)} \Gamma{\left(\frac{1}{2}-2\delta\right)} \Gamma{(2\delta)}.
}
Thus,
\eq{S_{TT}(0)|_{\mathrm{eq}}+2S_{0T}(0)|_{\mathrm{eq}} = 0.}
Taking into account that the Johnson-Nyquist noise of the lower edge $S_{00}(0)$ does not depend on the currents $I_n, I_s$, we get the expression \eqref{excess_noise} for the excess noise $\tilde{S}(0)$.

\section{Appendix E: Putting things together}

The expressions \eqref{noise_tun}-\eqref{j_s} for the ratio $X$ of the excess noise $\tilde{S}(0) = S_{TT}(0)+2S_{0T}(0)$ \eqref{excess_noise} and the tunneling rate $r = \left|I_T/I_s\right|$ \eqref{tun_general} can be straightforwardly obtained using the explicit expressions for $I_T$, $S_{TT}(0)$, $S_{0T}(0)$ in formulas \eqref{I_T_answer}, \eqref{S_TT_answer}, \eqref{S_0T_answer} respectively. We only changed the integration variable $\tau \rightarrow \pi T_s t$ and restored the fundamental constants: the elementary charge $e$, the Planck constant $h = 2 \pi \hbar$, and the Boltzmann constant $k_B$.

We remind the reader that in the main text of the paper we assumed the neutral mode chemical potentials of both edges $\mu^{(n, u)}, \mu^{(n, l)}$ to be zero. However, if needed, the neutral mode chemical potentials can be easily incorporated into the formulas \eqref{G_i}-\eqref{F_0T} by the substitution $Q_i j_s t \rightarrow \left(Q_i j_s - Q_i^{(n)} (\mu^{(n, u)}-\mu^{(n, l)})\right) t$. The neutral charges of the quasiparticles $Q_i^{(n)} = (\mathbf{g}_i)_2$ are given in the column $g_2$ of Table~\ref{tab:exc}.

\bibliography{FQHE_Noise_2_3}

\end{document}